# Super Phantoms: advanced models for testing medical imaging technologies

Srirang Manohar*,1, Ioannis Sechopoulos[1,2], Mark A. Anastasio[3], Lena Maier-Hein[4,5] and Rajiv (Raj) Gupta[6]

[1]*Multi-Modality Medical Imaging, Tech Med Centre, University of Twente, Enschede, The Netherlands*
[2]*Department of Medical Imaging, Radboud University Medical Center, Nijmegen, The Netherlands*
[3]*Department of Bioengineering, University of Illinois Urbana-Champaign, Urbana-Champaign USA.*
[4]*German Cancer Research Center, Heidelberg, Germany*
[5]*Ruprecht-Karls-University of Heidelberg, Heidelberg, Germany*
[6]*Department of Radiology, Massachusetts General Hospital, Boston, USA*
*s.manohar@utwente.nl

## Abstract

Phantoms are test objects used for initial testing and optimization of medical imaging techniques, but these rarely capture the complex properties of the tissue. Here we introduce super phantoms, that surpass standard phantoms being able to replicate complex anatomic and functional imaging properties of tissues and organs. These super phantoms can be computer models, inanimate physical objects, or ex-vivo organs. Testing on these super phantoms, will enable iterative improvements well before in-vivo studies, fostering innovation. We illustrate super phantom examples, address development challenges, and envision centralized facilities supporting multiple institutions in applying these models for medical advancements.

## Introduction

Human space flight stands as one of mankind's most remarkable technological achievements. Each step of this ambitious endeavor is accompanied by tremendous risks to both humans and machines. However, significant progress has been witnessed from Yuri Gagarin's pioneering spaceflight to the Apollo moon missions, culminating in the present era where crews spend extended periods aboard the International Space Station. One crucial element contributing to these achievements is the focus on rigorous testing and optimization on earth using specialized life-sized simulators and training modules. For instance, Neil Armstrong credited the Lunar Landing Training Vehicles [1] for making him feel "comfortably familiar" with the Eagle during the nerve-wracking descent to the lunar surface, where he maneuvered the craft to land in boulder-ridden territory while low on fuel [2].

We advocate for a comparable emphasis and commitment to rigorous laboratory testing during the development of advanced medical imaging technologies, prior to human testing. Medical imaging research and development are focused on achieving a future where imaging reveals quantitative imaging biomarkers of anatomy and physiology in three or more dimensions [3]. An imaging biomarker is defined as an objectively-measured characteristic extracted from an image used to evaluate normal and pathogenic biological processes. These quantitative biomarkers will be integral components in the decision-making process of precision medicine [4], with Artificial Intelligence (AI) underpinning every stage of the pipeline, from image reconstruction to integrative analytics and decision-making. However, in the face of this rapid pace of imaging technology advancements, the means and methodologies for testing and validating these technologies in controlled settings before embarking on human studies have not kept pace.

Laboratory testing of new imaging technologies does take place using phantoms. These are inanimate reference objects [5] used to assess the technical specifications of imaging instruments, including spatial, contrast, and temporal resolutions, as also to ascertain bias and precision. In certain cases, advanced phantoms are designed to simulate the anatomy of specific human body parts or organs targeted by the imaging technology. However, these phantoms often fail to replicate key aspects of the pathobiology being investigated, such as blood flow or perfusion. As a result, imaging phantom studies only provide an initial level of technical optimization. Any shortcomings or limitations in extracting qualitative and preferably quantitative imaging biomarkers are only discovered once human studies have begun which often occurs late in the development process. Human or in general *in vivo* testing cannot be conducted early during development and improvement stages for various reasons. Firstly, evaluation of accuracy or deviation of the imaging method from the gold standard or ground truth, is impossible since the *in-vivo* ground truth is unknown. Secondly, while precision or variability can be assessed through repeated imaging of human subjects, certain methods involving ionizing radiation or contrast agents make this impossible. Moreover, repeated imaging often fails to provide the necessary insights into error sources required for making essential improvements.

There is thus a major gap between testing on straightforward imaging phantoms and testing on highly complex human or animal tissues with their unknown ground truths. In this Perspective, we discuss the need for a new class of phantoms and their testing, which will be a bridge across this chasm. These are digital and physical phantoms that can replicate both the morphology and relevant aspects of physiology in a specific organ under both healthy and diseased conditions. These advanced phantoms would facilitate the early evaluation of an imaging method's ability to extract both anatomic and functional imaging biomarkers from acquired data. We call these models 'super phantoms' to signify that they occupy a class of physical and biological realism beyond standard imaging phantoms. We offer several examples of test objects that fit within our definition of super phantoms. We close with ways by which the community could arrive at a future paradigm where research into and application of super phantoms, is accepted as an integral component of research and development of imaging technologies.

## Super Phantoms

Phantoms are digital or physical test objects used for initial testing and optimization of new techniques, as well as for acceptance testing, accreditation, and quality control [5]. They have several attributes depending upon the task at hand, but in general are described as having tissue-like properties. This feature invariably only refers to the interaction properties of the phantom material with the energy beam of the imaging method used, such as x rays or ultrasound waves. They rarely capture the complex pathophysiological properties of the organ and their correlation with image features.

In contrast, we *introduce the definition of a "super phantom," as a computer model, an inanimate physical object, or an ex-vivo organ that replicates the most relevant anatomic and functional imaging biomarkers of the structure of interest under carefully controlled laboratory settings*. Examples of physiological and functional behavior which would produce the relevant imaging biomarkers are vasculature, flow, perfusion, diffusion, oxygen saturation and motion. Examples of structures of interest are organs, body parts, entire subjects (e.g. humans or small animals) and their various abnormalities. In the case of digital super phantoms, we can further define them as stochastic models capable of generating ensembles of phantoms. These ensembles reflect the statistical properties of biological variations present within the cohort to be imaged [6,7]. Additionally, we can define "blended digital-physical super phantoms" as a combination of both digital and physical elements, further expanding the capabilities of testing and optimization.

We illustrate with Figure 1 the need for super phantoms in two broad scenarios. The first scenario occurs in real-world clinical settings where an established imaging product exhibits inadequate performance in a specific application. To respond to this, improvements in the existing imaging technology or protocols may be recommended. In the second scenario, the clinical need may require a disruptive new imaging technology to be developed. In both scenarios, the conceptualization of the solution will require early testing with research prototypes on a surrogate of the human organ or body part.

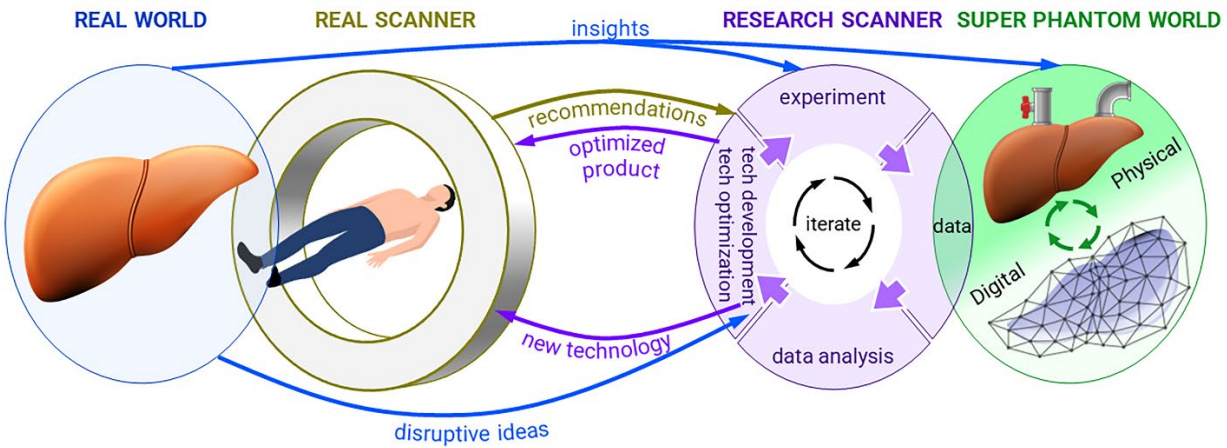

**Figure 1**: Mapping from real-world clinical setting to the research lab super phantom setting. The concept of "imaging super-phantoms" pertains to surrogates of tissues and organs that reproduce in vivo imaging features. These super phantoms are available in carefully-controlled laboratory settings to facilitate the assessment and refinement of emerging imaging technologies and methodologies, prior to human studies and clinical application.

During this early phase, digital phantoms can be employed for *in silico* testing and refinement of the technology and approaches. Following this, digital super phantoms would be used in simulations, for validation in controlled environments with reliable reference information. Subsequently, experimental studies using physical phantoms, and later super phantoms, should be conducted to evaluate the proposed imaging solution. These studies enable investigation of precision and accuracy, and the estimation of confidence intervals when extracting the required biomarkers. Recommendations for improvements and refinements can be iterated upon until an acceptable performance is achieved. Digital super phantoms will serve as valuable tools for conducting virtual imaging trials. These trials allow for the exploration of a vast range of potential solutions using objective measures of image quality [6,7]. Through this process, ensemble-averaged figures of merit can be computed, facilitating robust comparisons and refinement of the technology.

Once the technology solution has demonstrated the required level of performance on the super phantoms, the next step involves implementing it in *in vivo* studies. The studies would be on human subjects but also on small animals in the case of dedicated systems for fundamental studies. A significant advantage of conducting super phantom studies is that these models can represent the ground truth for various properties and imaging biomarkers. This not only enables reliable verification, and validation of new technologies but also provides data for training Machine Learning algorithms through imaging (both virtual and physical).

Examples of Super Phantoms

It is not our intention to provide an exhaustive overview of the phantom field nor the budding super phantom field, and will refer the reader to recent extensive reviews [9-14]. Here, we will briefly discuss three recently published phantoms that align with our definition of super phantoms.

1) A digital 4D CT breast imaging super phantom

With advancements in x-ray imaging using digital detectors, there has been a potential for functional imaging with high spatiotemporal resolution, particularly in applications such as cardiac and oncological imaging with the use of contrast agents. In the context of breast imaging, there has been a proposal for 4D imaging using dedicated breast CT, which aims to explore the perfusion properties of tumors for optimized treatment [14].

To understand the requirements and specifications of this new approach, Caballo et al. [15] developed a digital breast super phantom. This super phantom not only includes the morphological representation of breast tissues and tumors but also incorporates the wash-in and wash-out kinetics of iodinated contrast-containing blood in vasculature and simulated tumors. The morphology of the super phantom is based on segmented patient breast CT scans, which include four tissue types: skin, adipose tissue, fibroglandular tissue, and blood vessels. To create the digital breast super phantom, blood flow characteristics from breast MRI data were incorporated into the vessels, background parenchyma, and both benign and malignant tumors [15]. Other digital super phantoms for the breast have also been reported [16, 17].

2) A physical photoacoustic-ultrasound imaging super phantom

Photoacoustic imaging is an area of active research for imaging vasculature, hemodynamics, and blood oxygen saturation, which is particularly challenging in quantitative photoacoustics [18].

Dantuma et al. [19] recently presented a physical breast phantom specifically designed for quantitative hybrid photoacoustic and ultrasound imaging, which we classify as a super phantom. This phantom accurately represents the main tissues found in the female breast, including skin, fat, and fibroglandular tissue, using a custom polymer formulation with additives to achieve tissue-specific acoustic and optical interaction properties. The tissue-mimicking materials were meticulously cast layer-by-layer into shapes and sizes that resemble the breast, using 3D-printed molds based on a numerical breast model extracted from MRI images.

To simulate the larger vessels of the breast's vascular anatomy, the super phantom was designed to incorporate wall-less channels. Additionally, a flow circuit was developed to perfuse the channels with bovine blood at a controlled oxygen saturation level. This unique feature of the breast super phantom enables laboratory-based studies in photoacoustics, allowing for the validation and refinement of approaches to recover both qualitative and quantitative features that are sought after in *in-vivo* studies.

3) An ex-vivo 4D flow magnetic resonance imaging super phantom

4D phase-contrast magnetic resonance imaging holds great potential for accurate visualization and quantification of blood flow in three dimensions, along with the assessment of cardiac parameters [20]. However, conducting 4D flow MRI studies on live animals presents numerous challenges and limitations, including the lack of precise control over the disease process, ethical concerns, high costs, complex housing and maintenance requirements, and the need for anesthesia.

To address these limitations, an MRI-compatible platform was created centered around an isolated perfused beating pig heart, which meets the criteria of a super phantom. This platform was specifically designed to replicate *in-vivo* blood flow behavior, thereby generating physiological coronary flow and myocardial perfusion. The isolated heart model actively pumped blood in both ventricles, closely mimicking the natural filling of the coronary arteries and replicating physiological heart function with precise control over relevant physiological parameters. With this platform, researchers can investigate the impact of various surgical procedures on intra-cardiac flow and evaluate the performance of MRI sequences in different pathophysiological settings without the need for live animal experiments.

Figure 2 provides an overview of several advanced phantoms that have been recently published and can be classified as super phantoms according to our definition.

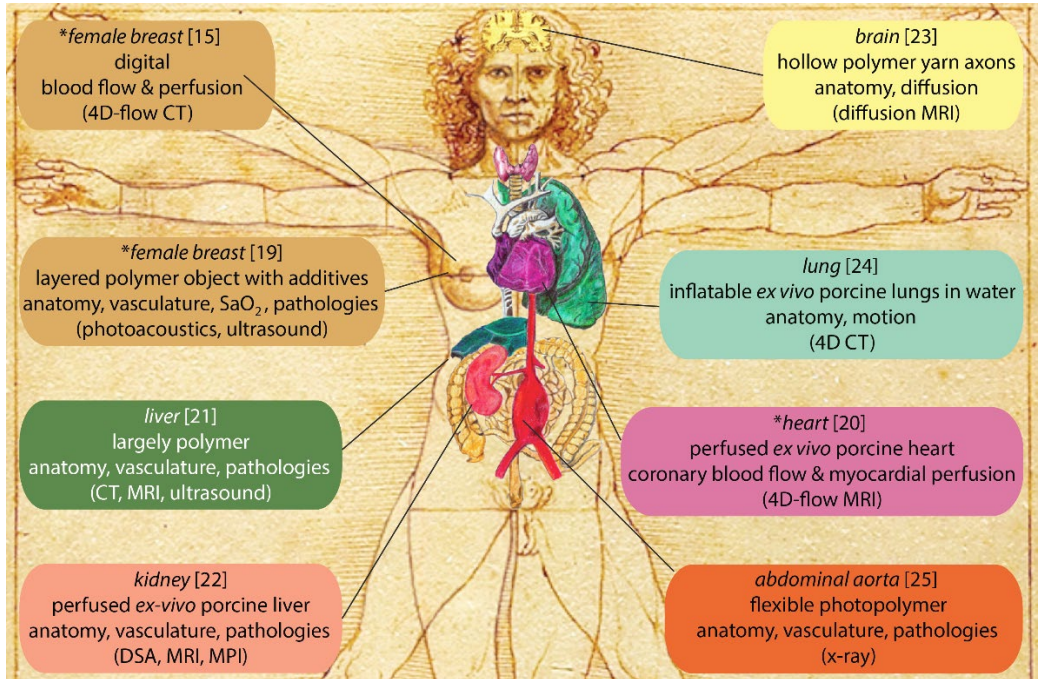

**Figure 2:** Examples of test-objects that we classify as super phantoms. The asterisks indicate those presented in text, reference numbers are provided in square brackets, and the imaging modality for which the super phantom is intended for is in round brackets. (Acronyms: 4D – four-dimensional, $SaO_2$ – oxygen saturation, CT –Computed Tomography, MRI – Magnetic Resonance Imaging, DSA – Digital Subtraction Angiography, MPI – Magnetic Particle Imaging.) The drawing of the body is inspired by the Vitruvian man created by Leonardo Da Vinci in 1490. The drawing was commissioned by the corresponding author, and he may use this in publications, presentations, web-sites and project proposals.

## Open challenges for super phantoms

It is important to acknowledge critical steps and considerations in the emerging field of super phantoms:

1. **Enhanced morphological and (patho)physiological representation:** Super phantoms as we define them, should aim on the one hand to provide an accurate and detailed representation of anatomical structures. This represents technical challenges in arriving at accurate variants to the extent necessary in terms of tissue composition and realistic geometry. On the other hand, super phantoms should also replicate physiological processes such as blood flow, perfusion, and tissue motion, which is technically. Further, there is also much to be learned about replicating various pathological conditions, such as tumors, vascular abnormalities, or tissue pathologies.

2. **Multi-modality integrated imaging:** These devices are implementations where the hardware of independent imaging modalities are physically integrated for complementarity leading to improved diagnostic and quantitative accuracy, and better understanding of disease-associated biological processes [26]. Examples of such hybrid imaging embodiments are SPECT (Single Photon Emission Tomography)/CT, PET (Positron Emission Tomography)/CT, and PET/MR imaging, which are available for clinical use and small-animal research [26]. Hybrid imaging throws up several challenges in cross-modality image reconstruction, image fusion and image interpretation, to new requirements for training and education of medical professionals. Further, validation of the clinical utility and accuracy of hybrid imaging approaches requires time-consuming and expensive studies. Laboratory studies will help surmount some of the challenges above except that there is a paucity of multi-modal phantoms and super phantoms. These will be challenging to develop as they should strive to possess the imaging biomarkers for multiple imaging modalities.
3. **Metrological underpinning**: Since super phantoms aim to represent the ground truth for certain quantitative imaging biomarkers, a metrological framework is essential [27]. Such a framework is to ensure that super phantom systems and their sub-systems such as Tissue-Mimicking Media (TMM)[28] are accurately characterized. These would be independent methods accepted as reference methods and calibrated to reference materials, and their properties labelled with the measurement uncertainty. Reporting of uncertainties will provide understanding of the reliability and limitations of these super phantoms. The measured parameters such as the size, density, flow etc. should be traceable to verifiable physical SI quantities. Such well characterized super phantoms would provide the ground truth to calibrate the specific imaging systems for which they were designed.
4. **Technological advancements:** There are vast opportunities for super phantom development to leverage technological advancements, such as 3D printing [29], novel materials, and advanced manufacturing techniques, to enhance the physical fidelity of phantoms. Incorporating smart materials, sensors, and actuators can enable real-time feedback and control, further improving realism and functionality.
5. **AI-based digital phantoms**: State-of-the-art frameworks for synthetic data generation typically leverage prior knowledge on tissue anatomy and functional tissue parameters for the generation of synthetic phantoms [30]. More recently, data-driven approaches are evolving as a promising alternative to the traditional model-based methods due to their ability to close the so-called domain gap between real and synthetic data [31, 32]. Such deep learning-based methods have the potential of producing large amounts of diverse and highly realistic digital phantoms [32]. A remaining challenge is the combination of such digital phantoms with physical phantoms in the spirit of blended digital-physical super phantoms. Initial steps have already been taken in the context of surgical training, in which traditional training (physical) boxes are combined with varying, potentially AI-generated, internal anatomy to practice surgical skills.
6. **Digital twins**: A related open challenge is the development of patient digital twins. While personalized 3D printing of individual organs for treatment planning and execution is becoming increasingly common, current methods often focus on only a specific aspect of patient-individual information, such as the 3D anatomy. Future work should be directed towards more holistic approaches, combining all the available information on a patient in one digital twin model, which could then serve a multitude of purposes, including outcome prediction, personalized therapy planning and even dry-runs of surgical intervention on the digital twin of a patient to be operated.
7. **Validation, standardization and collaboration:** Standardized protocols for super phantom fabrication and characterization as in 5. above, along with standardized protocols for their use in terms of image acquisition and analysis are essential for making super phantom reliable tools. This will facilitate reproducibility across different laboratories promoting inter-laboratory comparisons and enable the exchange of data with confidence [33,34]. EIBALL (European Institute for Biomedical Imaging

Research) and QIBA (Quantitative Imaging Biomarkers Alliance) are examples of initiatives focused on improving quantitative imaging biomarkers, where standardized ground truth phantoms [35,36] are crucial. Another initiative specific to the photoacoustic imaging modality is IPASC (International Photoacoustic Standardization Consortium) [37] one of whose tasks is to secure standardization of design, fabrication, and use of phantoms to validate and benchmark photoacoustic imaging systems. Similar efforts will be required for super phantoms as well, where collaboration among researchers is vital with sharing of phantom designs, fabrication and validation protocols, and imaging data.

8. **Improve awareness of the field:** The realization is growing that we need imaging on super phantoms, as an intermediate phase between imaging on standard phantoms and imaging on humans. However, more efforts should be made to improve awareness of super phantoms, and their roles in validating and assessing the latest generation of imaging technologies. This will provide an impulse to attract talented engineers to contribute to the field, and also help to draw interest from other academic domains such as material sciences and additive manufacturing.

## Next steps for super phantoms

To address the challenges and considerations above, it is necessary to establish super phantoms as an integral part of research and development in imaging technologies. We believe that this is possible with the following concrete steps:

### A. Establishment of Super Phantom Core Facilities

Phantom development is predominantly carried out by PhD students and Post-Docs within individual research groups. This fragmented and decentralized approach often leads to compromised instrumentation quality and redundancy due to limited individual budgets. To address this, we propose the establishment of a centralized shared facility that can cater to multiple institutions, equipped with high-quality instrumentation and dedicated personnel.

This facility would ideally be located at a university with a strong biomedical engineering program and established collaborations with academic hospitals, to house a range of equipment and supplies for the following:

1. **Fabrication and characterization of Tissue-Mimicking Media (TMM):** The facility would house a general chemistry laboratory, a 3D printing station [29], and a machine shop to prepare a variety of TMMs [28] using gels, elastomers, plastics, and resins. These TMMs would be doped with additives to replicate different tissue types. The characterization of TMMs would require instruments and tools to measure their interaction coefficients [38]. For example, an acoustic characterization scanning system [38, 39] could be employed to measure acoustic attenuation, sound speed, acoustic impedance, and non-linearity for ultrasound TMMs. These measurements are crucial for assessing the suitability of the TMMs and informing any necessary iterations or optimizations during the development process, and for establishing the ground truth values for the super phantom.
2. **Super phantom platform development:** TMMs would be assembled to construct the body part or organ of interest. This stage would involve incorporating physiological features and behaviors into the super phantom, such as circulation, perfusion, and motion. The facility would facilitate the integration of these elements to create the super phantoms.
3. **Validation and experimentation:** The complete super phantom would be imaged using the specific modality for which it was designed. The facility would require access to imaging equipment, similar to that used in clinical settings, to replicate the real-world operational environment. Close collaboration with radiologists and clinicians would be essential for the analysis and interpretation of the acquired images. The primary goals would be to validate the super phantom's ability to meet the quantitative functional requirements and to conduct studies using research variants of imaging equipment to implement improvements in technology or protocols.
4. **Computing:** The facility would need high-performance computing resources for generation of digital super phantoms as well as for conducting *in silico* experiments and trials. Furthermore, numerical procedures for objective and quantitative assessment of both digital and physical super phantoms would be implemented using these tools.

### B. Infrastructure for Collaboration and Standardization

The facility would serve as a central hub for visiting research groups from the imaging and biomedical engineering communities, fostering collaborative efforts in the field of super phantoms. It would also provide the necessary infrastructure for intramural imaging studies, promoting a paradigm of collaborative experiments. Customized super phantoms could be disseminated within the community for measurements and investigations, including inter-laboratory studies.

Setting up and operating such a facility would entail significant costs and necessitate substantial initial investment through multi-institutional support and various programs offered by national and international funding agencies. To ensure sustainability, the facility's running costs would be covered by project grants, with internal and external users budgeting for their use of the facility.

To promote knowledge sharing and collaboration, protocols for fabrication and characterization measurements on physical phantoms, along with the resulting data, would be made accessible to the community through databases. Digital super phantoms would be open source, enabling imaging technology developers worldwide to readily access them for virtual imaging studies. This imaging data would be invaluable for qualitative and quantitative testing of image reconstruction, processing, and analysis algorithms and for training machine learning models. To host the data effectively, appropriate metadata would require to be recorded with domain-specific ontology for semantic search and data integration.

Online collaboration environments would facilitate consensus building on various aspects related to super phantoms, aiming to establish standards. The facility would play a vital role in developing and maintaining standard protocols for super phantoms, serving as a repository of knowledge and expertise in both physical and digital variants of these phantoms.

## Concluding Remarks

We firmly believe that super phantoms, capable of mimicking the physical and functional characteristics of tissues and organs in both healthy and diseased states, will play a crucial role in the research and development of new imaging technologies. These phantoms offer the opportunity for comprehensive assessments and iterative optimization of imaging modalities at lower Technology Readiness Levels (TRLs) [40], enabling improvements in their chances of success before human studies. The outcomes of studies using these sophisticated phantoms may also provide the first evidence of readiness for regulatory boards when assessing applications for subsequent clinical studies and CE (conformité européenne) certification [41].

In order to integrate super phantom work into the research and development of imaging technologies, it is essential to establish shared facilities housing the necessary infrastructure for designing, developing, characterizing, and validating super phantoms. Such a super-phantom facility could serve as a hub for researchers from within and outside the hosting institution, fostering improved rigor in studies and facilitating the free flow of ideas for continuous enhancements of the super phantoms and the studies conducted with them.

With growing awareness of the nature and significance of super phantoms and the establishment of such facilities, talented biomedical engineers will be attracted to engage in the research, development, and utilization of these phantoms. The field is replete with numerous fundamental and applied questions and challenges, offering ample opportunities for interdisciplinary research. For instance, we still have much to discover regarding the relevant interaction properties of imaging modalities (optical, acoustic, x-ray, etc.) in various pathologies, which is essential for developing super phantoms that accurately reflect clinical reality. The level of realism required in super phantoms and how to define and measure this realism also represent important and rich areas of exploration. Naturally, technical challenges abound, such as accurately mimicking and measuring pathophysiological flow and perfusion in different organs.

The role of industry will be important in the future for dissemination and use of super phantoms The acquisition and use of these phantoms for assessing and validating expensive advanced imaging systems hold significant commercial appeal. As awareness of super phantoms grows, this nascent sector is expected to steadily expand, encouraging increased industry participation and direct investment in the future.

While our Perspective has centered on super phantoms for imaging applications, these tools can play a roles in planning therapeutic interventions such as radiation therapy [42], thermal ablations [43], and surgery [44]. The principles and insights discussed in our work can serve as a foundation for application of super phantoms in these areas. More research and development will be necessary to simulate the biological and physical responses of tissue subjected to therapeutic interventions.

In summary, we strongly believe that the field of super phantoms will become increasingly important, significantly accelerating progress, fostering breakthroughs, and expediting the clinical translation of new and effective imaging modalities and approaches.

## Data Availability Statement
No datasets were generated or analyzed during the current study.

## Acknowledgements
SM acknowledges members of the Imaging and Diagnostics Domain at the University of Twente for discussions. SM thanks the Strategic Impulse Program Personalized Health of the Tech Med Centre, University of Twente for support. SM acknowledges the help of Mr. Arthur Veugelers (University of Twente) in preparing Figure 1, and Mr. Lew Sołowiej (independent artistic collaborator ) in preparing Figure 2. We thank Prof. E.G.E. (Elisabeth) de Vries of the University Medical Centre Groningen, and Dr. A.H. (Anneline) Jonker of the University of Twente, for their review of an early version of the manuscript.

## Author contributions
SM, IS, MAA, LMH and RG discussed early ideas leading to the concepts presented in the Perspective. SM wrote the draft of the manuscript with inputs from IS, MAA, LMH and RG. SM conceived and developed the figures with help as in Acknowledgements. All authors read the draft, provided feedback and contributed with improvements in text to lead to the submitted version. All authors contributed to the rebuttal and to the final accepted version.

# Competing interests
All authors state they have no competing interests.

**establish global standards for photoacoustic imaging predominantly by defining test objects (phantoms), establishing test methods and providing open datasets.**

*************